\documentclass[%
 aip,
 jmp,%
 amsmath,amssymb,
preprint,%
]{revtex4-1}
\usepackage[dvipdfm,bookmarks=true,colorlinks=true,citecolor=blue,linkcolor=blue,urlcolor=blue,linkbordercolor={1 1 1},citebordercolor={1 1 1},urlbordercolor={1 1 1}]{hyperref}%
\usepackage{graphicx}
\usepackage{dcolumn}
\usepackage{bm}
\usepackage{epstopdf}
\begin{document}

\title{Reduced hierarchical equations of motion in real and imaginary time: Correlated initial states and thermodynamic quantities}

\author{Yoshitaka TANIMURA}
 \email{tanimura@kuchem.kyoto-u.ac.jp.}
\affiliation{Freiburg Institute for Advanced Studies, Albert-Ludwigs-Universit\"at Freiburg, Albertstr. 19, Freiburg 79104, Germany}
\affiliation{Department of Chemistry, Graduate School of Science, Kyoto University,
Kyoto 606-8502, Japan}
\date{\today}

\begin{abstract}
For a system strongly coupled to a heat bath, the quantum coherence of the system and the heat bath plays an important role in the system dynamics. This is particularly true in the case of non-Markovian noise. We rigorously investigate the influence of system-bath coherence by deriving the reduced hierarchal equations of motion (HEOM), not only in real time, but also in imaginary time, which represents an inverse temperature. It is shown that the HEOM in real time obtained when we include the system-bath coherence of the initial thermal equilibrium state possess the same form as those obtained from a factorized initial state. We find that the difference in behavior of systems treated in these two manners results from the difference in initial conditions of the HEOM elements, which are defined in path integral form. We also derive HEOM along the imaginary time path to obtain the thermal equilibrium state of a system strongly coupled to a non-Markovian bath. Then, we show that the steady state hierarchy elements calculated from the real-time HEOM can be expressed in terms of the hierarchy elements calculated from the imaginary-time HEOM. Moreover, we find that the imaginary-time HEOM allow us to evaluate a number of thermodynamic variables, including the free energy, entropy, internal energy, heat capacity, and susceptibility. The expectation values of the system energy and system-bath interaction energy in the thermal equilibrium state are also evaluated. 

\end{abstract}

\pacs{Valid PACS appear here}
                         

\maketitle

\section{INTRODUCTION}
Quantum open systems have been a subject of fundamental interest for many years. 
Problems in this category include those of understanding how the irreversibility of time appears in system dynamics, why macroscopic systems can be treated with classical mechanics instead of quantum mechanics, how wave functions collapse as a result of measurements done with macroscopic instruments, and why and how quantum systems approach a thermal equilibrium state through interaction with their environments.\cite{CaldeiraPhysica83,LeggettRMP87, Weiss08, KuboToda85}
Theories of quantum open systems have also been used to construct models of practical interest, in particular to account for line shapes in EPR, NMR\cite{Redfield65, Kubo69}  and laser spectra,\cite{Mukamel95} to evaluate chemical reaction rates\cite{HanggiRMP90} and electron and charge transfer rates\cite{MarcusRMP90, Khun95} in chemical physics, and to explore the lifetimes of quantum entanglement states in quantum information theory.\cite{Breuer02}

The phenomena mentioned above arise from the unavoidable interaction of a system with its environment. In the quantum mechanical case, dissipative systems are often modeled as main systems coupled to heat-bath degrees of freedom at finite temperature. This coupling gives rise to thermal fluctuations and dissipation that drive the systems toward the thermal equilibrium state. The heat-bath degrees of freedom are then reduced using such methods as the projection operator method and the path integral method.

The projection operator approach is effective if the interaction between the system and the bath is weak. If one further assumes that the correlation time of the noise arising from the interaction with the bath is very short (the Markovian assumption), equations of motion for the density matrix elements can be derived, and these can be solved numerically. The most commonly used equations of this kind are the quantum master equations\cite{KuboToda85,Weiss08} and the Redfield equation.\cite{Redfield65}
It has been proven, however, that quantum master equations and the Redfield equation do not satisfy the necessary positivity condition.\cite{Davies76,Gorini78,Spohn80,Dumcke79}  Careful analyses addressing this problem have been carried out by several researchers.\cite{Pechukas94,Romero04}
As a method to preserve positivity, the rotating wave approximation (RWA), which modifies the interaction between the system and the heat  bath, has been applied.\cite{Frigerio81,Frigerio85} However, this may alter the thermal equilibrium state as well as the dynamics of the original total Hamiltonian.
Then, building on these results, it was shown that the violation of the positivity condition results from the Markovian assumption. Specifically, it was found that even if the dissipation process is Markovian, the fluctuation process may not be, because it must satisfy the fluctuation-dissipation theorem.\cite{TanimuraJPSJ06}

The time convolution-less (TCL) master equation has a wider range of applicability than the quantum master equations and Redfield equation, because it allows the system-bath interaction to be non-perturbative and fluctuation and dissipation to be non-Markovian.\cite{Shibata77,Shibata79} In order for a non-perturbative treatment to be possible, however, the system Hamiltonian and the bath interactions of the TCL equation must commute.\cite{ShiTCL09,Ban10} Thus, the TCL equation cannot be used to treat many important problems involving molecules, atoms, and spins driven by a time-dependent laser or magnetic field. In addition, because of the factorized nature of the system-bath interaction, in the case of a non-commuting excitation, the TCL equation cannot be used to calculate nonlinear response functions of the system operator involved in the optical multidimensional spectrum.\cite{IshizakiCP08,TanimuraACR09}

Path integral Monte Carlo simulations do not have the limitations of any of the approaches discussed above, and for this reason, they are capable of incorporating imaginary path integrals and unfactorized initial conditions more easily,  
but this approach is computationally heavy, because the number of paths to be evaluated grows rapidly with time, while sampling fails due to the phase cancellation of wave functions.\cite {EggerMak94,Makri95,CaoVoth96} Much effort has been made to extend the applicability of this method.\cite{Makri96,Thorwart00,Makri07,JadhaoMakri08, Dattani12B, Dattani13} Because this approach can easily incorporate the semi-classical approximation in the bath, it may have an advantage in the study of polyatomic systems treated in multi-dimensional coordinates, but applications to this point incorporating full quantum dynamics have been limited to relatively small systems.

Many of the above-mentioned limitations can be overcome with the hierarchal equations of motion (HEOM) for the reduced density matrix, which are derived  by differentiating the reduced density matrix elements defined by path integrals.\cite{TanimuraJPSJ06} This approach was introduced to investigate the connection between the phenomenological stochastic Liouville equation and the dynamical Hamiltonian theory, and was originally limited to the case in which the spectral distribution function takes the Drude form (i.e., the Ohmic form with a Lorentzian cutoff) and the bath temperature is high.\cite{Tanimura89A} However, with the inclusion of low temperature corrections terms, this temperature limitation has been eliminated.\cite{TanimuraPRA90,IshizakiJPSJ05,Xu05,Yan06} 
In addition, with the extension of the dimension of the hierarchy, this approach is capable of treating a great variety of spectral distribution functions.\cite{TanimuraMukamelJPSJ94,TanakaJPSJ09,TanakaJCP10,TanimruaJCP12,Nori12,KramerFMO,YanBO12,Shi14}
This formalism is valuable because it can treat not only strong system-bath coupling but also quantum coherence between the system and bath, which is essential to calculate nonlinear response functions. 
The system-bath coherence becomes particularly important if the bath interaction is regarded as non-Markovian, as was found from nonlinear optical measurements in the late 1980s, when laser technology reached the femto-second time scale, which is much shorter than the noise correlation time of environmental molecules.\cite{Mukamel95} 
The HEOM approach has been used to study such problems, which include multi-dimensional spectroscopy.\cite{Tanimura89B,Tanimura89C,TanimuraMaruyama97, KatoTanimura04,
IshizakiTanimuraJCP06,IshizakiTanimuraJPCA07,SakuraiJPC11,DijkstraDNA10,ShiYan2D,Shi11FMO,KramerNJP12FMO, KramerJPC2013} Recently, it was shown that system-bath coherence also plays an important role in calculations of quantum measures involving concurrence\cite{Breuer02} and non-Markovianity\cite{BreuerPRL09} under  multiple external perturbations.\cite{DijkstraPRL10,DijkstraPhil12,DijkstraJPSJ12} 

Because the HEOM approach is computationally heavy, a variety of methods have been developed to study dissipative dynamics in realistic situations. \cite{Shi09,YanPade10A,YanPade10B,Aspru11,GPU11,Schuten12,Cao2013,Shi2013} It has been applied to the study of multi-dimensional vibrational spectroscopies,\cite{KatoTanimura04,IshizakiTanimuraJCP06,IshizakiTanimuraJPCA07,SakuraiJPC11} photosynthetic antenna systems,\cite{Shi11FMO,KramerNJP12FMO,KramerJPC2013, Ishizaki09,Schuten09,Schuten11,DijkstraNJP12} fermion systems, \cite{JinYanFermi07,JinYanFermi08,YanKondo09} quantum ratchets,\cite{KatoJPCB13} resonant tunneling diodes,\cite{SakuraiJPSJ13,SakuraiNJP14} and dissociation of tightly bounded electron-hole pairs.\cite{YaoYaoJCP14}

While the applicability of the HEOM approach continues to expand, the basic nature of the hierarchy elements has not been thoroughly explored. The purpose of this paper is to investigate the role of correlated initial equilibrium states in the HEOM formalism. Until this time, the HEOM have been derived by assuming a factorized initial state, $\hat \rho_{tot} =\exp[-\beta \hat H_A] \exp[-\beta \hat H_B]$, at inverse temperature $\beta$, where $\hat H_A$ and $\hat H_B$ are the system and bath Hamiltonians, respectively, while the true thermal equilibrium state of the system is given by $\hat \rho_{tot} =\exp[-\beta (\hat H_A +\hat H_I +\hat H_B )]$, where $\hat H_I$ is the system-bath interaction. The difference between the factorized and correlated initial states becomes large for strong $\hat H_I$.
Analysis based on an analytic solution of a Brownian oscillator system indicates that even if we start from a factorized initial state, the system reaches the true equilibrium state, $tr_B \{\exp[-\beta (\hat H_A +\hat H_I +\hat H_B )]\}$, through transient phenomena arising from the factorized initial state, for example phenomena known as initial sweeping.\cite{Haake85,GrabertPR88,Grabert97}
With the HEOM approach, we have run the HEOM program until all of the hierarchy elements reach the steady state and then used these elements as the initial conditions of the correlated thermal equilibrium state. The accuracy of this method has been confirmed by analyzing multi-dimensional spectra obtained with it.\cite{IshizakiTanimuraJCP06}
Nevertheless, it would be interesting to derive the HEOM starting from a correlated initial thermal state in order to obtain an analytically derived expression for the system-bath coherence in the HEOM formalism. Moreover, with a simple generalization, we can also derive the HEOM in imaginary time, which corresponds to the inverse temperature. We show that the imaginary-time HEOM is convenient for obtaining correlated thermal equilibrium states and the thermodynamic variables of the reduced system.

The organization of the paper is as follows. In Sec. II we present a model Hamiltonian and its influence functional with correlated initial states.  In Sec. III, we derive the HEOM from the density matrix elements using the influence functional formalism with the correlated initial states given in Sec. II. In Sec. IV, we derive the imaginary-time HEOM, which is convenient for evaluating correlated thermal equilibrium states and the thermodynamic quantities of the system.
In Sec. V, to confirm the validity and numerical efficiency of our approach, we report the results of numerical integrations of the HEOM carried out over real time and imaginary time for a spin-boson system and compare their results. Thermodynamic variables and expectation values for the spin-boson system are also calculated as a demonstration.  Section VI is devoted to concluding remarks.

\section{INFLUENCE FUNCTIONAL WITH CORRELATED INITIAL STATES}
We consider a situation in which the system interacts with a heat bath that gives rise to dissipation and fluctuation in the system. To illustrate this, let us consider a Hamiltonian expressed as
\begin{align}
\hat H_{tot} &= \hat H_A + \hat H_I + \hat H_B,
\label{eq:SpinBoson}
\end{align}
where $\hat H_A \equiv H_A (\hat a^+,\,\hat a^-  )$ is the Hamiltonian of the system, denoted by A, defined by the creation and annihilation operators  $\hat a^{+}$ and $\hat a^{-}$. The bath degrees of freedom are treated as an ensemble of harmonic oscillators, 
\begin{align}
\hat H_B = \sum\limits_{j} {\left( \frac{\hat p_j^2 }{2m_j } + \frac{1}{2}m_j \omega _j^2 \hat x_j^2 \right) },
\label{eq:bath}
\end{align}
with the momentum, position, mass, and frequency of the $j$th bath oscillator given by $\hat{p}_{j}$, $\hat{x}_{j}$, $m_{j}$ and $\omega_{j}$, respectively.
The system-bath interaction is given by
\begin{align}
\hat H_I = - \hat V( \hat a^+,\,\hat a^-   )\sum\limits_j {\alpha_j \hat x_j},
\label{eq:interaction}
\end{align}
where  $\hat V(\hat a^+,\,\hat a^- )$ is the system part of the interaction, and $\alpha_j$ is the coupling constant between the system and the $j$th oscillator.

The heat bath can be characterized by the spectral distribution function, defined by
\begin{eqnarray}
  J(\omega) \equiv \sum_{j }\frac{\hbar
 \alpha_{j}^2}{2m_{j}\omega_{j}} \delta(\omega-\omega_{j}),
\label{eq:J_wgeneral}
\end{eqnarray}
and the inverse temperature, $\beta \equiv 1/k_{\mathrm{B}}T$, where $k_\mathrm{B}$ is the Boltzmann constant. Note that if $\hat a^{+}$ and $\hat a^{-}$ respectively represent the creation and annihilation operators of spin states, the above Hamiltonian is the spin-boson Hamiltonian,~\cite{LeggettRMP87,Weiss08} which has been studied with various approaches.\cite{Redfield65, Kubo69,Mukamel95,EggerMak94,Makri95,CaoVoth96} 

Now, let us introduce the fermion coherent state $\lvert \phi \rangle$, which satisfies $\hat a^{-} \lvert \phi \rangle = \phi \lvert \phi \rangle$ and $\langle \phi \rvert \hat a ^{+}  =  {\phi}^{\dag} \langle \phi \rvert$, where $\phi$ and ${\phi}^{\dag}$ are Grassmann numbers (G-numbers).
\cite{Tanimura89A,TanimuraPRA90,IshizakiJPSJ05}
Note that here we consider a two-level system, but extension to a multi-level system is also straightforward.\cite{TanakaJPSJ09,TanakaJCP10} In practice, we can treat a G-number system in the same manner as a c-number system, as long as we maintain the time order of the operators in the integral. In the path integral representation, the time propagator of the wave function (the Feynman propagator) for the total system is expressed as
\begin{eqnarray}
  G_{tot}(\phi^{\dag},{\bold x},\phi_0, {\bold x}_0;\,t) &&= \frac1{N}\int_{\phi(0)=\phi_0}^{\phi(t)=\phi}  {D[\phi ^{\dag}(\tau) \phi (\tau)]} \int_{{\bold x}(0)={\bold x}_0}^{{\bold x}(t)={\bold x}}  D[{\bold x}(\tau)] \nonumber \\
&&\times {\rm e}^{\frac{\rm i}{\hbar } S_A[\phi ^{\dag},\,\phi ;\,t] +\frac{i}{\hbar}  \int_0^t d\tau \left[ \frac12{m \dot {\bold x}^2(\tau)}- \frac12 m \omega^2 {\bold x}^2(\tau) +  V (\tau){\bold x} (\tau) \right]},  
\label{Gtot}
\end{eqnarray}
where $N$ is the normalization constant, 
$\int D[\phi ^{\dag}(\tau)\phi (\tau)]$ represents a functional integral over a set of Grassmann variables, and $\int D[{\bold x}(\tau)]\equiv \Pi_j \int D[x_j(\tau)]$ denotes path integrals over the bath oscillator coordinates, with $m \dot {\bold x}^2\equiv \sum_j m_j{\dot x}_j^2$, $m \omega^2 {\bold x}^2\equiv \sum_j m_j \omega_j^2 x_j^2$, and $V(\tau){\bold x}(\tau)\equiv V \left( \phi ^{\dag}(\tau),\,\phi(\tau) \right) \sum_j \alpha_j x_j(\tau)$. Here, the action for the system's Hamiltonian, $\hat H_A (\hat a^+,\,\hat a^- )$, is denoted by $S_A[{\phi} ^{\dag},\phi; t]=\int^t_{0} d\tau L_A({\phi}^{\dag},{\phi})$, with the Lagrangian
\begin{eqnarray}
L_A({\phi}^{\dag},{\phi})= {i \hbar}\phi^{\dag}\dot {\phi} -H_A({\phi}^{\dag},{\phi}).
\end{eqnarray}
The thermal equilibrium state can also be expressed in the path integral representation by making the replacement $i \tau/\hbar \to \tau'$ in Eq.\eqref{Gtot}. We thereby obtain
\begin{eqnarray}
  \rho_{tot}^{eq}(\phi_0,{\bold x}_0, {\phi'}_0^{\dag}, {\bold x}'_0;\,\beta \hbar) &&= \frac{1}{Z_{tot}}\int_{\bar \phi(0)=\phi'_0}^{\bar \phi(\beta\hbar)=\phi_0}  {D[\bar \phi ^{\dag}(\tau' ) \bar \phi (\tau' )]} \int_{\bar {\bold x}(0)={\bold x}'_0}^{\bar {\bold x}(\beta\hbar)={\bold x}_0}  D[\bar {\bold x}(\tau')] \nonumber \\
&&\times {\rm e}^{-\frac{\rm 1}{\hbar } \bar S_A[\bar \phi ^{\dag},\,\bar \phi ;\,\beta \hbar] -\frac{1}{\hbar}  \int_0^{\beta \hbar} d\tau' \left[ \frac12{m \dot {\bar {\bold x}}^2(\tau')}+ \frac12 m \omega^2 \bar {\bold x}^2(\tau') -  {\bar V} (\tau') \bar {\bold x} (\tau') \right]},  
\label{GEqtot}
\end{eqnarray}
where $Z_{tot}$ is the normalization constant, the G-numbers $\{ {\bar \phi} ^{\dag}, \bar \phi\}$ form the coherent representations of the operators $\{ \hat a^+,\,\hat a^- \}$ for the equilibrium distribution, $ \bar S_A[\bar \phi ^{\dag}, \bar \phi; \beta \hbar]$ is the Euclid action of the system obtained from  $S_A[{\phi} ^{\dag},\phi; t]$ through the replacement $i t/\hbar  \to \beta$, and ${\bar V}(\tau')\equiv V\left( \bar \phi^{\dag}(\tau'),\,\bar \phi(\tau') \right)$. The total density matrix elements is then given by
 \begin{eqnarray}
  \rho_{tot} (\phi ^{\dag},{\bold x}, \,\phi', {\bold x}';\,t) &&= 
\int {\int {d\phi_0 ^{\dag}d\phi_0 } \int {\int { d\phi_0 'd{\phi_0 '}^{\dag}}} } \int d{\bold x}_0 \int d{\bold x}_0' \nonumber \\
&&\times G_{tot} (\phi^{\dag},{\bold x},\phi_0, {\bold x}_0,\,t) \rho_{tot}^{eq}(\phi_0,{\bold x}_0, {\phi_0'}^{\dag}, {\bold x}_0';\,\beta \hbar) G_{tot}^{\dag} (\phi',{\bold x}',{\phi_0'}^{\dag}, {\bold x}_0',\,t).
\label{rhotot0} 
\end{eqnarray}
The heat-bath degrees of freedom can be eliminated by integrating over the bath coordinates as
$\rho (\phi ^{\dag},\,\phi ';\,t) = \int d{\bold x} \rho_{tot} (\phi ^{\dag},{\bold x}, \,\phi', {\bold x};\,t)$. The reduced density operator is then 
expressed in the coherent representation of the G-numbers as\cite{Tanimura89A,TanimuraPRA90,IshizakiJPSJ05}
\begin{eqnarray}
\hat \rho  (t) =  \int {\int {d\phi ^{\dag}d\phi } \int {\int {d{\phi '}^{\dag}} d\phi '} } \left| \phi  \right\rangle \rho (\phi ^{\dag},\,\phi ';\,t)\left\langle {\phi '} \right|,
\end{eqnarray}
where
\begin{eqnarray}
  \rho (\phi ^{\dag},\,\phi ';\,\,t) &&= \frac{1}{Z_{tot}}\int_{\phi(0)=\phi_0}^{\phi(t)=\phi}  {D[\phi ^{\dag}(\tau ) \phi (\tau )]} \int_{\bar \phi (0)=\phi'_0}^{\bar \phi (\beta \hbar)=\phi_0}  D[{\bar \phi}^{\dag}(\tau' ) \bar \phi (\tau' )]  
\int_{\phi'(0)=\phi_0'}^{\phi'(t)=\phi'} 
 {D[{\phi '}^{\dag}( \tau )\phi '(\tau )]  
 }  \nonumber \\  
 && \times \bar \rho_0^{eq}[\bar \phi ^\dag  ,\bar \phi; \beta \hbar ] {\rm e}^{\frac{\rm i}{\hbar } S_A[\phi ^{\dag},\,\phi ;\,t]} 
 F[\bold V; t, \beta\hbar] {\rm e}^{ - \frac{\rm i}
{\hbar }S_A^{\dag}[{\phi '}^{\dag},\,\phi ';\,t]}.
\label{rhotot1}
\end{eqnarray}
Here, $\rho_0^{eq}[\bar \phi ^\dag  ,\bar \phi; \beta \hbar]$ is the thermal equilibrium distribution of the system $A$ itself, defined by the Euclid action, $\bar S_A[\bar \phi ^{\dag}, \bar \phi; \beta \hbar]$,
and $F[\bold V; t, \beta\hbar]$ with $\bold V \equiv \{ V( {\phi'}^{\dag},\,\phi' ),  V( \bar \phi ^{\dag},\,\bar \phi ), V( \phi ^{\dag},\,\phi ) \}$ is the influence functional for correlated initial states.\cite{GrabertPR88} Employing the counter path illustrated in Fig. 1, we can express the influence functional  in the path integral representation as (see Appendix A)
\begin{figure}
\begin{center}
\scalebox{0.5}{\includegraphics{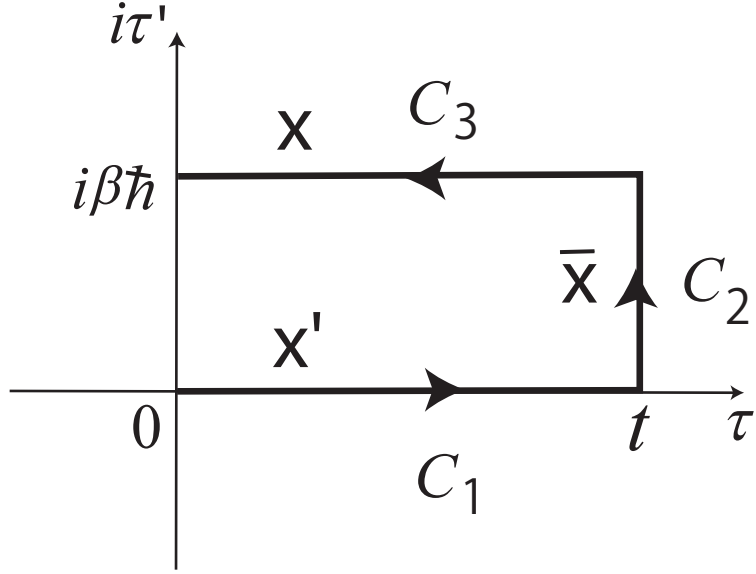}}
\end{center} 
\caption{\label{counterpath} The counter path for the influence functional given in Eq.\eqref{eq:F_x}.} 
\end{figure}
\begin{eqnarray}
F[\tilde V_C ; t, \beta\hbar] = \int d\bold x  \int_{{\bold x}'(t)={\bold x}}^{{\bold x}(t)={\bold x}}  D[\tilde {\bold x}(s)]
e^{-\frac{i}{\hbar} \int_C ds\left[ \frac12{m \dot {\tilde {\bold x}}^2(s)}- \frac12 m \omega^2\tilde {\bold x}^2(s) + \tilde V_C(s)\tilde {\bold x} (s) \right]},  
\label{eq:F_x}
\end{eqnarray}
where $\int D[\tilde {\bold x}(s)]\equiv \Pi_j \int D[x_j(\tau)] \int D[\bar x_j(\tau')] \int D[x_j'(\tau)]$, with the contour path defined by
\begin{eqnarray}
\int_C ds = \int_0^t ds + \int_{t}^{t+i \beta \hbar} ds + \int_{t+i \beta \hbar}^{i \beta \hbar} ds,
\label{eq:counterpath}
\end{eqnarray}
and
\begin{eqnarray}
\tilde {\bold x} (s) = 
 \left\{  
  	\begin{array}{c}
    		{\bold x}'(t-s) \,\, \text{on} \,\, C_1 \\ 
    		\bar {\bold x} (-i(s-t)) \,\, \text{on} \,\,  C_2 \\ 
		{\bold x} (t-s+i\hbar \beta) \,\, \text{on} \,\, C_3 \\ 
  	\end{array}
 \right.,~
\tilde V_C(s) = 
 \left\{  
  	\begin{array}{c}
    		V' (t-s) \,\, \text{on} \,\, C_1\\
    		{\bar V}(-i(s-t)) \,\, \text{on} \,\, C_2     \\
		V (t-s+i\hbar \beta) \,\, \text{on} \,\, C_3	 \\	
  	  	\end{array}
 \right.
\label{eq:counterV}
\end{eqnarray}
with  $V(\tau)\equiv V( {\phi ^{\dag}(\tau),\,\phi(\tau) } ) $, $V'(\tau) \equiv V( {{\phi '}(\tau),\,{\phi'}^{\dag}(\tau) } )$ and  $\bar V(\tau') \equiv V({\bar \phi} ^{\dag}(\tau'), \bar \phi(\tau'))$, respectively.
The path integral used here to derive the HEOM is expressed in terms of an influence functional. The calculation of the influence functional for a heat bath consisting of harmonic oscillators is analogous to that of the generating functional for a Brownian oscillator system if we regard the system operator in the system-bath interaction $\hat V$ as an external force acting on the bath. \cite{Tanimura93,OkumuraPRE96,TanimuraOkumuraJCP97,SuzukiTanimuraPRE99} Then, the influence functional can be calculated analytically and is found to be $F[\tilde V_C; t, \beta\hbar]=\exp\left\{{\tilde \Phi}[\tilde V_C ; t, \beta\hbar]\right\}$, where the influence phase is expressed as (see Appendix A) 
\begin{eqnarray} 
{\tilde \Phi}[\tilde V_C ; t, \beta\hbar] =  \frac1{\hbar^2} 
\int_C ds'' \int_{C'} ds' \tilde V_C(s'') \tilde V_{C'}(s') L (s'' - s' ).
\label{eq:PhitA}
\end{eqnarray}
Here, $C'$ represents the counter path for $s'$ that follows $s''$ along $C$ under the condition $s''> s'$ 
and 
\begin{eqnarray}
L(t+i\tau)
=\int_0^{\infty}  d\omega J(\omega)\frac{1}{\sinh\left( \frac{\beta \hbar\omega}2 \right)}
&&\left[ \cosh\left( \frac{\beta \hbar\omega}2- \omega \tau \right)\cos(\omega t) \right.  \nonumber \\
&& \left .+i \sinh\left( \frac{\beta \hbar\omega}2- \omega \tau \right)\sin(\omega t) \right].
\label{eq:Ltitau}
\end{eqnarray}
After dividing the contour of the integral in Eq.\eqref{eq:PhitA} into $C_1$, $C_2$ and $C_3$, we have (see Appendix B)
\begin{eqnarray}
{\tilde \Phi}[\bold V; t, \beta\hbar] &=& - \frac1{\hbar^2} \int_0^t dt'' \int_0^{t''} dt' 
V^\times(t'') \left[-i L_1  (t''-t') V^\circ(t') +L_2(t''-t') V^\times(t') \right] \nonumber \\
&&+ \frac{i}{\hbar^2}  \int_0^t dt'' \int_0^{\beta \hbar} d \tau' V^\times(t'') 
L (t'' + i \tau' ) \bar V (\tau') 
\nonumber \\ &&
+ \frac1{\hbar^2} \int_0^{\beta \hbar}  d \tau'' \int_0^{\tau''}  d \tau' \bar V (\tau'')  \bar V (\tau') 
\bar L(\tau'' -  \tau '),
\label{eq:PhitB}
\end{eqnarray}
where we define $L(t)\equiv iL_1(t)+L_2(t)$, $\bar L(\tau) \equiv L(i \tau)$ and
\begin{align}
V^{\times}(t) &\equiv V(t) - V'(t),  \nonumber \\
V^{\circ}(t) &\equiv V(t) + V'(t).
\label{eq:VtimesVcirc}
\end{align}
The functionals $V^{\times}(t)$ and $V^{\circ}(t)$ represent the commutator and anticommutator of $\hat V$. This form of the influence functional has been used to analytically study quantum Brownian systems.\cite{GrabertPR88} 

The first term in Eq.\eqref{eq:PhitB} represents a commonly used influence functional derived from the factorized initial conditions. \cite{CaldeiraPhysica83,LeggettRMP87, Weiss08}  
The collective bath oscillator coordinate, $\hat{X}\equiv \sum_{j} \alpha_{j}\hat{x}_{j}$, is regarded as a driving force for the system through the interaction $-\hat{V}\hat{X}$.
The time-dependent kernels are then represented by $i L_1 (t) \equiv \langle [\hat X(t), \hat X]\rangle$ 
and $L_2(t) \equiv \langle \hat X(t) \hat X+\hat X \hat X(t) \rangle /2$, respectively, where 
$\hat{X}(t)$ is the Heisenberg representation of the operator $\hat{X}$.
\cite{TanimuraJPSJ06} 
The function $L_2(t)$ is analogous to the classical correlation function of the bath induced noise $X(t)$ and corresponds to the fluctuations. The dissipation corresponding to $\bar C (t)=-\int dt L_1(t)$ is related to $L_2 (t)$ through the quantum version of the fluctuation-dissipation theorem,
$L_2[\omega] = \hbar \omega \coth(\beta \hbar \omega/2) \bar C [\omega]/2$,
which insures that the system evolves toward the thermal equilibrium state for finite temperatures.\cite{KuboToda85} 
The second term in Eq.\eqref{eq:PhitB} consists of the cross-terms between the real-time and imaginary-time integrals that describe the correlation between the initial equilibrium state and the dynamical state at time $t$. This term represents the contribution of the correlated initial conditions and can be regarded as the non-Markovian effects with respect to both real and imaginary times. 
The last term describes the influence of the heat bath on the thermal equilibrium state of the system. In the following sections, in order to derive the imaginary-time HOEM, we consider the full equilibrium state of the system, $\rho^{eq}[\bar \phi ^\dag  ,\bar \phi; \beta \hbar]$, by including the last term in $\rho_0^{eq}[\bar \phi ^\dag  ,\bar \phi; \beta \hbar]$.

For $0<\tau<\beta\hbar$, we have\cite{GrabertPR88,OkumuraPRE96,TanimuraOkumuraJCP97}
\begin{eqnarray}
\frac{\cosh\left( \frac{\beta \hbar\omega}2- \omega \tau  \right)}{\sinh\left( \frac{\beta \hbar\omega}2 \right)} = \frac{2}{\beta \hbar} \sum_{k=-\infty}^{\infty} \frac{\omega e^{ i \nu_k \tau}}{\omega^2+\nu_k^2},
\end{eqnarray}
\begin{eqnarray}
\frac{\sinh\left( \frac{\beta \hbar\omega}2- \omega \tau  \right)}{\sinh\left( \frac{\beta \hbar\omega}2 \right)} = \frac{2}{\beta \hbar} \sum_{k=-\infty}^{\infty} \frac{- i \nu_k e^{ i \nu_k \tau}}{\omega^2+\nu_k^2},
\end{eqnarray}
where $\nu_k \equiv 2 k \pi /\hbar \beta$ are the Matsubara frequencies. We thus have
\begin{eqnarray}
L(t+i \tau) =   \frac{2}{\beta\hbar}\int_{0}^{\infty}  d\omega J(\omega)  
\left[\frac1{\omega} + \sum_{k=1}^{\infty} \frac{2\omega}{\nu_k^2+\omega^2}  \cos (\nu_k \tau)  \right] \cos(\omega t) \nonumber \\
 +i  \frac{2}{\beta\hbar} \int_{0}^{\infty}  d\omega J(\omega)  
 \sum_{k=1}^{\infty} \frac{2\nu_k}{\nu_k^2+\omega^2}  \sin (\nu_k \tau)  \sin(\omega t) ,
\label{eq:LFourier}
\end{eqnarray}
and
\begin{eqnarray}
\bar L(\tau) &&=   \frac{2}{\beta\hbar} \int_0^{\infty}  d\omega J(\omega) \left[\frac1{\omega} + \sum_{k=1}^{\infty} \frac{2\omega}{\nu_k^2+\omega^2} \cos(\nu_k \tau) \right] .
\label{eq:Lbardef} 
\end{eqnarray}

For the case $\tau=0$, we use the definition given in Eq.\eqref{eq:Ltitau} to obtain 
\begin{equation}
L_1(t) = \int_0^{\infty}  d\omega J(\omega)\sin(\omega t),
\end{equation}
\begin{eqnarray}
L_2(t)=  \frac{2}{\beta\hbar} \int_{0}^{\infty}  d\omega J(\omega)  \left[\frac1{\omega} + \sum_{k=1}^{\infty} \frac{2\omega}{\nu_k^2+\omega^2} \right] \cos(\omega t).
\end{eqnarray}

\section{REDUCED HIERARCHAL EQUATIONS OF MOTION IN REAL TIME}

We assume that the spectral density $J(\omega)$ has an Ohmic form
with a Lorentzian cutoff and write\cite{TanimuraJPSJ06}
\begin{equation}
 J(\omega) = \frac{\hbar \eta}{\pi}\frac{\gamma^2\omega}{\gamma^2+\omega^2},
\label{JDrude}
\end{equation}
where the constant $\gamma$ represents the width of the spectral distribution of the collective bath modes
and is the reciprocal of the correlation time of the noise induced by the bath. 
The parameter $\eta$ is the system-bath coupling strength, which represents the magnitude of damping.

With Eq.\eqref{JDrude} for $0<\tau<\beta\hbar$, we obtain
\begin{eqnarray}
L(t+i \tau) = &&
\left\{ c_0''  +\sum_{k=1}^{\infty}    \textcolor{red}{  \left[c_k''\cos (\nu_k \tau ) + i c_k'  \sin (\nu_k \tau ) \right] } \right\} e^{-\gamma t} \nonumber \\
&&  +  \sum_{k=1}^{\infty}  \textcolor{red}{ c_k' \left[\cos (\nu_k \tau ) -i \sin (\nu_k \tau ) \right] }
 e^{-\nu_k t}, 
\label{eq:Ltitau2}
\end{eqnarray}
where
\begin{eqnarray}
c_k'  = - \frac{2\eta \gamma ^2 }{\beta}
\frac{{\nu_k}}{{\gamma^2  - \nu_k^2 }},
\label{eq:c_k}
\end{eqnarray}
\begin{eqnarray}
c_k''  =  \frac{2\eta \gamma ^2 }{\beta }
 {\frac{{ \gamma }}{{\gamma^2  - \nu_k^2 }}},
\label{eq:b_k}
\end{eqnarray}
and $c_0'' = {\eta \gamma }/{\beta}$.  
At $t=0$, the above equation reduces to
\begin{eqnarray}
\bar L(\tau) =  \sum_{k=0}^{\infty}  \bar c_k \cos (\nu_k \tau),
\label{eq:LbarD}
\end{eqnarray}
where $\nu_0\equiv 0$, $\bar c_0=c_0''$, and  $\bar c_k \equiv c_k'  + c_k''$ for $1 \le k$, while at $\tau=0$, we have
\begin{eqnarray}
L_1(t) = \frac{ \textcolor{red}{  \hbar }
\eta \gamma^2}{2}
{\rm e}^{ - \gamma \left| t \right|},
\label{eq:L_1barGM}
\end{eqnarray}
and
\begin{eqnarray}
L_2(t) &&= c_0' e ^{ - \gamma |t|}  + \sum\limits_{\textcolor{red}{  k}
 = 1}^{\infty}  {c_k' } e ^{ - \nu _k |t|} 
\nonumber \\
&&\approx c_0' e ^{ - \gamma |t|}  + \sum\limits_{ \textcolor{red}{  k}
 = 1}^{K}  {c_k' } e ^{ - \nu _k |t|} 
+  \delta(t) \sum\limits_{ \textcolor{red}{  k}
 = K+1}^{\infty} \frac{c_k'}{\nu_k},
\label{eq:L_2GMDef2}
\end{eqnarray}
with $c_0'  = c_0'' + \sum_{k=1}^\infty c_k''=\hbar \eta\gamma^2 \cot (\beta\hbar\gamma/2)/2$.
Here, we choose $K$ so as to satisfy $\nu_k=2\pi K/(\beta\hbar) \gg \omega_c$, where $\omega_c$ represents the characteristic frequency of the system. Under this condition we can apply the approximation $\nu_k{\rm e}^{-\nu_k |t|}\simeq \delta( t ) \quad ({\rm for} \ \ j \geq  K+1)$ with negligible error at the desired temperature, $1/\beta$.

We define the equilibrium distribution function of the system under the influence of the heat bath through the replacement of the last term of ${\tilde \Phi}[\bold V; t, \beta\hbar]$ (as expressed in Eq.\eqref{eq:PhitB}) appearing $F[\bold V; t, \beta\hbar]$ with $\rho_0^{eq}[\bar \phi ^\dag  ,\bar \phi; \beta \hbar]$. We then obtain
\begin{eqnarray}
\rho^{eq}[\bar \phi ^\dag  ,\bar \phi; \beta \hbar]=\frac{1}{Z_B}
 {\rm exp}&&\left[- \frac1{\hbar} \bar {S_A}[\bar \phi ^
 {\dag}, \bar \phi; \beta \hbar]  \right.\nonumber \\
&&\left. + \sum_{k=0}^{\infty} \frac{\bar c_k}{\hbar^2}  
\int_{0}^{ \beta \hbar } d \tau''  \int_{0}^{\tau''} d  \tau'  \bar V(\tau'') \bar V( \tau') 
\cos \left( \nu_k (\tau''- \tau')\right) \right]. 
\label{eq:PCI}
\end{eqnarray}
The influence functional $ F[\bold V; t, \beta\hbar]$ is redefined through this replacement as 
\begin{eqnarray}
F_{CI}[\bold V; t, \beta\hbar] &&=  \exp \left[
 \left( -\frac{i}{\hbar}  \right)^2 \int_{0}^t dt'' \operatorname{e} ^{ - \gamma t'' }   
V^{\times}(t'' ) \left( \int_{0}^{t''} dt'  
  \operatorname{e} ^{  \gamma  t' } \Theta (t' ) 
-i \bar \Theta  (\beta \hbar)  \right)  
 \right] \nonumber \\
&&\times \exp \left[ \left( -\frac{i}{\hbar}  \right)^2
\int_{0}^t dt''\sum\limits_{k= 1}^{K} \operatorname{e} ^{ - \nu_k t'' }  
V^{\times}(t'' )\left( \int_{0}^{t''} dt' 
  \operatorname{e} ^{  \nu_k  t' } \Psi_k (t' ) 
-i \bar \Psi_k   (\beta \hbar) \right)  
 \right]  \nonumber \\
&&\times \exp \left[ -  \int_{0}^t dt'' \Xi(t'') \right],
\label{eq:influenceFcoh}
\end{eqnarray}
where
\begin{eqnarray}
\Theta (t) \equiv c_0'
V^{\times}(t)
 -  \frac{{\rm i}{\hbar \eta \gamma ^2 }}{2} V^{\circ}(t),
\label{eq:Theta}
\end{eqnarray}
\begin{eqnarray}
\bar \Theta (\beta \hbar) \equiv 
\int_{0}^{\beta\hbar} d \tau'  \bar V(\tau')
\left\{c_0''+\sum_{k=1}^{\infty}   \textcolor{red}{  \left[c_k'' \cos (\nu_k \tau' ) + i c_k'  \sin (\nu_k \tau' ) \right] }
\right\},
\label{eq:Thetabar}
\end{eqnarray}
and for $k \ge 1$,
\begin{eqnarray}
\Psi_k (t ) \equiv c_k' V^{\times}(t),
\label{eq:Psi_k}
\end{eqnarray}
\begin{eqnarray}
\bar \Psi_k  (\beta \hbar) \equiv 
\int_{0}^{\beta\hbar} d \tau'  \bar V(\tau')
\textcolor{red}{ c_k' \left[\cos (\nu_k \tau' ) -i \sin (\nu_k \tau' ) \right],}
\label{eq:Psi_kbar}
\end{eqnarray}
and
\begin{eqnarray}
\Xi(t) \equiv - \sum_{k=K+1}^{\infty}  \textcolor{red}{ \frac{1}{\nu_k}}
  V^ \times (t)  {\Psi}_k (t).
\label{eq:Xi}
\end{eqnarray}
Note that in the high temperature limit, $\beta\hbar\gamma \ll 1$,
the noise correlation function reduces to $L_2(t) \propto {\rm e}^{ - \gamma \left| t \right|}$.
This indicates that the heat bath oscillators interact with the system in the form of Gaussian Markovian noise.\cite{Tanimura89A}

The equations of motion for the reduced density operator can be derived by evaluating 
the time derivative of the wavefunctions on the left-hand and right-hand sides and the influence functional.\cite{TanimuraJPSJ06,Tanimura89A, TanimuraPRA90, IshizakiJPSJ05, Xu05,Yan06,TanimuraMukamelJPSJ94,TanakaJPSJ09,TanakaJCP10,TanimruaJCP12} 
If we consider the auxiliary matrix defined by 
\begin{eqnarray}
\rho_{j_1,\dots,j_K}^{(n)}({{\phi}^{\dag}},\phi';t)&&= 
\frac{1}{Z_{tot}}\int_{\phi(0)=\phi_0}^{\phi(t)=\phi}  {D[\phi ^{\dag}(\tau ) \phi (\tau )]} \int_{\bar \phi (0)=\phi'_0}^{\bar \phi (\beta \hbar)=\phi_0}  D[{\bar \phi}^{\dag}(\tau ') \bar \phi (\tau' )]  
\int_{\phi'(0)=\phi_0'}^{\phi'(t)=\phi'} 
 {D[{\phi '}^{\dag}(\tau )\phi '(\tau )]  
 } \nonumber \\  
 && \times \bar \rho^{eq} [\bar \phi ^\dag  ,\bar \phi; \beta \hbar ] {\rm e}^{\frac{\rm i}{\hbar } S_A[\phi ^{\dag},\,\phi ;\,t]} 
F_{j_1, \cdots,j_K}^{(n)}[ \bold V; t, \beta\hbar] {\rm e}^{ - \frac{\rm i}
{\hbar }S_A^{\dag}[{\phi '}^{\dag},\,\phi ';\,t]}
\label{eq:rhon_jk}
\end{eqnarray}
where
\begin{eqnarray}
F_{j_1, \cdots,j_K}^{(n)}[ \bold V; t, \beta\hbar] &&= 
 \left\{ \textcolor{red}{ -\frac{i}{\hbar }}
\int_{0}^{t} dt' 
  \operatorname{e} ^{ -\gamma (t-t') }\Theta (t' ) 
-i \operatorname{e} ^{ -\gamma t }\bar \Theta (\beta \hbar) \right\} ^{n} \nonumber \\
&& \times \prod_{k=1}^K 
\left\{ \textcolor{red}{ -\frac{i}{\hbar }}
\int_{0}^{t} dt' 
  \operatorname{e} ^{ -\nu_k  (t-t')}\Psi_k (t' ) 
-i \operatorname{e} ^{ -\nu_k t} \bar \Psi_k  (\beta \hbar)\right\} ^{j_k} 
\nonumber \\&&\times 
F_{CI}[\bold V; t, \beta\hbar]
\label{auxFV}
\end{eqnarray}
for nonnegative integers $n,j_1,\dots,j_K$. 
Among the $\hat{\rho}_{j_1,\dots,j_K}^{(n)}(t)$, only $\hat{\rho}^{(0)}_{0,\dots,0}(t)=\hat{\rho}(t)$ has a physical meaning, and the others are introduced for computational purposes. 
Differentiating $\rho^{(n)}_{j_1,\dots,j_K}({{\phi}^{\dag}},\phi';t)$ with respect to $t$, we obtain the following hierarchy of equations in operator form:
\begin{eqnarray}
	\frac{\partial}{\partial t}
	\hat{\rho}_{j_1,\dots,j_K}^{(n)}(t)
	&&=-\left[
		\frac{i}{\hbar} \hat{ H }_A^{\times} + n \gamma+
		\sum_{k=1}^K j_k\nu_k
		+\hat{\Xi}
	\right]
	\hat{\rho}_{j_1, \dots,j_K}^{(n)}(t) \notag\\
     &&-\frac{i}{\hbar}\hat V^{\times}
	\hat{\rho}_{j_1, \dots,j_K}^{(n+1)}(t) - \frac{i}{\hbar} \sum_{k=1}^K \hat V^{\times} 
	\hat{\rho}_{j_1, \dots,j_k+1,\dots,j_K}^{(n)}(t)	\notag\\
		&&-  \frac{i n}{\hbar} \hat{\Theta}
	\hat{\rho}_{j_1,\dots,j_K}^{(n-1)} (t)	- \sum_{k=1}^K 
\frac{i j_k}{\hbar}	 \hat{\Psi}_k
	\hat{\rho}_{j_1,\dots,j_k-1,\dots,j_K}^{(n)}(t),
\label{eq:HEOM}
\end{eqnarray}
where $\hat{H}^{\times}$ is the Liouvillian of $\hat H_A$, and
the relaxation operators $\hat{\Theta}$ and $\hat{\Psi}_k$ 
 are obtained through the replacement $V^\times(t)\to\hat{V}^\times$ and $V^\circ(t)\to\hat{V}^\circ$ in Eqs.\eqref{eq:Theta} and \eqref{eq:Psi_k}, where
$\hat{\mathcal{O}}^\times\hat{f}	\equiv	\hat{\mathcal{O}} \hat{f}-\hat{f}\hat{\mathcal{O}}$
and 
$\hat{\mathcal{O}}^\circ\hat{f}	\equiv \hat{\mathcal{O}}\hat{f}+\hat{f}\hat{\mathcal{O}}$
for any operand operator $\mathcal{\hat{O}}$ and $\hat{f}$,
  and
\begin{eqnarray}
  \hat{\Xi} \equiv \left\{-\frac{\eta}{\beta} \left[1 - \frac{\beta\hbar\gamma}{2}\cot\left( \frac{\beta\hbar\gamma}{2} \right) \right]
   + \sum_{k=1}^K  \frac{ c_k'}{\nu_k}  \right\} 
	\hat{V}^\times\hat{V}^\times.
\label{HEOMT3}
\end{eqnarray}
The above expression is identical to the HEOM with a factorized initial state and can be truncated in the same manner as in the factorized case for large $N\equiv n+\Sigma_{k=1}^K j_k \gg \omega_c/min(\gamma, {\color{red}{\nu_1}}) $, where $\omega_c$ is the characteristic frequency of the system.\cite{TanimuraJPSJ06,IshizakiJPSJ05} If we add the counter term to the Hamiltonian \eqref{eq:SpinBoson}, we have an additional term in Eq.\eqref{HEOMT3}.\cite{IshizakiTanimuraJCP06}

While the terms from the correlated initial state $\bar \Theta$ and $\bar \Psi_k$ do not appear in Eq.\eqref{eq:HEOM}, they define the hierarchy elements for the correlated initial equilibrium state. To demonstrate this point, we consider the initial states of the density operators, obtained by setting $t=0$ in Eqs.\eqref{eq:rhon_jk} and \eqref{auxFV}: 
\begin{eqnarray}
\rho_{j_1,\dots,j_K}^{(n)}({{\phi}^{\dag}},\phi';0) = \frac{1}{Z_{A}} \int_{\bar \phi (0)=\phi'_0}^{\bar \phi (\beta \hbar)=\phi_0}  D[{\bar \phi}^{\dag}(\tau ) \bar \phi (\tau )]    
 \left(-i \bar \Theta  (\beta \hbar) \right)^{n}   \prod_{k=1}^K 
\left(  -i  \bar \Psi_k  (\beta \hbar) \right)^{j_k} 
 \bar \rho [\bar \phi ^\dag  ,\bar \phi; \beta \hbar ].   \nonumber \\
\label{eq:correlatedHEOM}
\end{eqnarray}
Here, ${Z_{A}}=Z_{tot}/Z_{B}$ and
\begin{eqnarray}
  \bar \rho [\bar \phi ^\dag  ,\bar \phi; \tau ] &&=
 {\rm exp}\left[ - \frac1{\hbar} \bar {S_A}[\bar \phi ^
 {\dag}, \bar \phi; \tau] \right] 
  \nonumber \\
&& \times  {\rm exp}\left[ 
 \sum_{k=0}^{\infty} \frac{\bar c_k}{\hbar^2}  
\int_{0}^{\tau} d \tau''  \int_{0}^{\tau''} d  \tau'  \bar V(\tau'') \bar V( \tau') 
\cos (\nu_k \tau'')\cos (\nu_k \tau')  \right] \nonumber \\
&& \times  {\rm exp}\left[ 
 \sum_{k=1}^{\infty}  \frac{\bar c_k}{\hbar^2}  
\int_{0}^{\tau} d \tau''  \int_{0}^{\tau''} d  \tau'  \bar V(\tau'') \bar V( \tau') 
\sin (\nu_k \tau'')\sin (\nu_k \tau')  \right],
\end{eqnarray}
and we have $\rho^{eq}[\bar \phi ^\dag  ,\bar \phi; \beta \hbar]= Z_{B} \bar \rho [\bar \phi ^\dag  ,\bar \phi; \beta \hbar]$.
This defines the correlated equilibrium initial conditions of Eq.\eqref{eq:HEOM}. In the next section, we derive the equations of motion to evaluate these hierarchy elements. 

\section{REDUCED HIERARCHAL EQUATIONS OF MOTION IN IMAGINARY TIME: CORRELATED THERMAL EQUILIBRIUM STATE}

The thermal equilibrium state $\bar \rho[\bar \phi ^\dag  ,\bar \phi; \tau ]$ at time $t=0$ and inverse temperature $\tau$ can be obtained by considering the imaginary-time derivative of the reduced density matrix elements given in Eq.(\ref{eq:PCI}).
This is expressed as
\begin{eqnarray}
\frac{\partial}{\partial \tau} 
\hat {\bar \rho}_{k^1,\dots,k^{m}}^{\,[m:l]} (\tau)
&& = - \hat H_A \hat {\bar \rho}_{k^1,\dots,k^{m}}^{\,[m:l]}  (\tau)  
 +  \frac{1}{\hbar}
\sum\limits_{k^{m+1} =0}^{K} \bar c_{k^{m+1}} \cos(\nu_{k^{m+1}} \tau)   
\hat V \hat {\bar \rho}_{k^1,\dots,k^{m+1}}^{\,[m+1:l]}  (\tau)
 \nonumber  \\
&&  +   \frac{1}{\hbar}
\sum\limits_{k^{m+1} =0}^{K} \bar c_{k^{m+1}} \sin(\nu_{k^{m+1}} \tau)   
\hat V \hat {\bar \rho}_{k^1,\dots,k^{m+1}}^{\,[m+1:l+1]}  (\tau) \nonumber  \\
&& 
+  \frac{1}{\hbar}\sum\limits_{h =1}^{m-l}\cos(\nu_{k^h} \tau)  \hat V \hat {\bar \rho}_{k^1,\dots,k^{h-1},k^{h+1},\dots,k^{m}}^{\,[m-1:l]}  (\tau)
 \nonumber  \\
&& +  \frac{1}{\hbar}\sum\limits_{h =m-l+1}^{{\color{red}{m}}}\sin(\nu_{k^h} \tau) \hat V \hat {\bar \rho}_{k^1,\dots,k^{h-1},k^{h+1},\dots,k^{m}}^{\,[m-1:l-1]}  (\tau),
\label{eq:ImHEOM}
\end{eqnarray}
where $\hat {\bar \rho}_{k^1,\dots,k^{m}}^{\,[m:l]}(\tau)$ is the density operator defined in path integral form as
\begin{eqnarray}
\bar \rho_{k^1,\dots,k^{m}}^{\,[m:l]} (\phi _0^\dag  ,\phi '_0; \tau) &&= \int_{\bar \phi (0)=\phi'_0}^{\bar \phi (\tau)=\phi_0} 
D[\bar \phi ^{\dag}(\tau ) \bar \phi (\tau )]
  \prod_{g=1}^{m-l} \left( \int_0^{\tau}  d\tau_g  \cos(\nu_{k^g}\tau_g)   \bar V(\tau_g) \right) \nonumber \\
&&\times\prod_{g'=m-l+1}^{m} \left( \int_0^{\tau}  d\tau_{g'}  \sin(\nu_{k^{g'}}\tau_{g'})  \bar V(\tau_{g'}) \right) \bar \rho[\bar \phi ^\dag  ,\bar \phi; \tau ].
\label{eq:ImHEOMelm}
\end{eqnarray}
Note that the first product in Eq.\eqref{eq:ImHEOMelm}  contains $(m-l)$ factors, and the second contains $l$ factors. Thus, the expression there is $(m-l)$th order in $\left( \int d\tau \cos(\nu_k \tau) \right)$
and $l$th order in $\left( \int d\tau \sin(\nu_k \tau) \right)$.  
Also, $ \hat {\bar \rho}_{k^1,\dots,k^{h-1},k^{h+1},\dots,k^{m}}^{\,[m-1,l]}  (\tau)$ and 
$ \hat {\bar \rho}_{k^1,\dots,k^{h-1},k^{h+1},\dots,k^{m}}^{\,[m-1,l-1]}  (\tau)$ in Eq.\eqref{eq:ImHEOM} denote the hierarchy elements defined by Eq.\eqref{eq:ImHEOMelm} without the index $k^h$ for $0 \le h \le m$. 
Note that any exchange of suffixes $k^i$ and $k^j$ in Eq.\eqref{eq:ImHEOMelm} that merely results in the permutation of two cosine factors or two sine factors leaves the total integral unchanged, while one that results in the arguments of a sine and cosine being exchanged will generally cause the total integral to change. To truncate the hierarchy equations, we choose 
some large value of $K'\equiv m$  
and set the elements at $(m+1)$th order to 0. We thus obtain a closed set of equations
up to $m$th order. 

To illustrate the structure of the hierarchy given in Eq.\eqref{eq:ImHEOM}, here we write out the equations up to second order. The hierarchy starts from the zeroth-order equation, which is that for the thermal equilibrium state density matrix:
\begin{eqnarray}
\frac{\partial}{\partial \tau} 
\hat {\bar \rho}^{\,[0:0]} (\tau)
&& = - \hat H_A \hat {\bar \rho}^{\,[0:0]}  (\tau)  
\nonumber  \\
&& +  \frac{1}{\hbar}
\sum\limits_{k^{1} =0}^{K'} \bar c_{k^{1}} \cos(\nu_{k^{1}} \tau)   
\hat V \hat {\bar \rho}_{k^{1}}^{\,[1:0]}  (\tau)  +  \frac{1}{\hbar}
\sum\limits_{k^{1} =0}^{K'} \bar c_{k^{1}} \sin(\nu_{k^{1}} \tau)   
\hat V \hat {\bar \rho}_{k^1}^{\,[1:1]}  (\tau).
\end{eqnarray}
Then, the first order consists of two equations,
\begin{eqnarray}
\frac{\partial}{\partial \tau} 
\hat {\bar \rho}_{k^1}^{\,[1:0]} (\tau)
&& = - \hat H_A \hat {\bar \rho}_{k^1}^{\,[1:0]}  (\tau)  +  \frac{1}{\hbar}\cos(\nu_{k^1} \tau) \hat V \hat {\bar \rho}^{\,[0:0]}  (\tau)
\nonumber  \\
&& +  \frac{1}{\hbar}
\sum\limits_{k^{2} =0}^{K'} \bar c_{k^{2}} \cos(\nu_{k^{2}} \tau)   
\hat V \hat {\bar \rho}_{k^1,k^{2}}^{\,[2:0]}  (\tau)
+  \frac{1}{\hbar}
\sum\limits_{k^{2} =0}^{K'} \bar c_{k^{2}} \sin(\nu_{k^{2}} \tau)   
\hat V \hat {\bar \rho}_{k^1,k^{2}}^{\,[2:1]}  (\tau),
\end{eqnarray}
\begin{eqnarray}
\frac{\partial}{\partial \tau} 
\hat {\bar \rho}_{k^1}^{\,[1:1]} (\tau)
&& = - \hat H_A \hat {\bar \rho}_{k^1}^{\,[1:1]}  (\tau)  +  \frac{1}{\hbar}\sin(\nu_{k^1} \tau) \hat V \hat {\bar \rho}^{\,[0:0]}  (\tau)
\nonumber  \\
&& +  \frac{1}{\hbar}
\sum\limits_{k^{2} =0}^{K'} \bar c_{k^{2}} \cos(\nu_{k^{2}} \tau)   
\hat V \hat {\bar \rho}_{k^1,k^{2}}^{\,[2:1]}  (\tau)+ \frac{1}{\hbar}
\sum\limits_{k^{2} =0}^{K'} \bar c_{k^{2}} \sin(\nu_{k^{2}} \tau)   
\hat V \hat {\bar \rho}_{k^1,k^{2}}^{\,[2:2]}  (\tau),
\end{eqnarray}
and the second order consists of three equations,
\begin{eqnarray}
\frac{\partial}{\partial \tau} 
\hat {\bar \rho}_{k^1,k^2}^{\,[2:0]} (\tau)
&& = - \hat H_A \hat {\bar \rho}_{k^1,k^{2}}^{\,[2:0]}  (\tau)  
+  \frac{1}{\hbar} \cos(\nu_{k^1} \tau) \hat V \hat {\bar \rho}_{k^{2}}^{\,[1:0]}  (\tau)+ \frac{1}{\hbar} \cos(\nu_{k^2} \tau) \hat V \hat {\bar \rho}_{k^{1}}^{\,[1:0]}  (\tau)
 \nonumber \\ 
&& + \frac{1}{\hbar}
\sum\limits_{k^{3} =0}^{K'} \bar c_{k^{3}} \cos(\nu_{k^{3}} \tau)   
\hat V \hat {\bar \rho}_{k^1,k^2,k^3}^{\,[3:0]}  (\tau)  +  \frac{1}{\hbar} \sum\limits_{k^{3} =0}^{K'} \bar c_{k^3} \sin(\nu_{k^{3}} \tau)   
\hat V \hat {\bar \rho}_{k^1,k^2,k^{3}}^{\,[3:1]}  (\tau),
\end{eqnarray}
\begin{eqnarray}
\frac{\partial}{\partial \tau} 
\hat {\bar \rho}_{k^1,k^2}^{\,[2:1]} (\tau)
&& = - \hat H_A \hat {\bar \rho}_{k^1,k^{2}}^{\,[2:1]}  (\tau)  
+  \frac{1}{\hbar} \cos(\nu_{k^1} \tau) \hat V \hat {\bar \rho}_{k^{2}}^{\,[1:1]}  (\tau)
+ \frac{1}{\hbar} \sin(\nu_{k^2} \tau) \hat V \hat {\bar \rho}_{k^1}^{\,[1:0]}  (\tau)  \nonumber \\ 
&&+  \frac{1}{\hbar}
\sum\limits_{k^{3} =0}^{K'} \bar c_{k^{3}} \cos(\nu_{k^{3}} \tau)   
\hat V \hat {\bar \rho}_{k^1,k^2,k^3}^{\,[3:1]}  (\tau)  +  \frac{1}{\hbar}\sum\limits_{k^{3} =0}^{K'} \bar c_{k^3} \sin(\nu_{k^{3}} \tau)   
\hat V \hat {\bar \rho}_{k^1,k^2,k^{3}}^{\,[3:2]}  (\tau),
\end{eqnarray}
\begin{eqnarray}
\frac{\partial}{\partial \tau} 
\hat {\bar \rho}_{k^1,k^2}^{\,[2:2]} (\tau)
&& = - \hat H_A \hat {\bar \rho}_{k^1,k^{2}}^{\,[2:2]}  (\tau)  
+  \frac{1}{\hbar} \sin(\nu_{k^1} \tau) \hat V \hat {\bar \rho}_{k^{2}}^{\,[1:1]}  (\tau)+ \frac{1}{\hbar} \sin(\nu_{k^2} \tau)  \hat V \hat {\bar \rho}_{k^{1}}^{\,[1:1]}  (\tau)
 \nonumber \\ 
&&+  \frac{1}{\hbar}
\sum\limits_{k^{3} =0}^{K'} \bar c_{k^{3}} \cos(\nu_{k^{3}} \tau)   
\hat V \hat {\bar \rho}_{k^1,k^2,k^3}^{\,[3:2]}  (\tau)  +  \frac{1}{\hbar} \sum\limits_{k^{3} =0}^{K'} \bar c_{k^3} \sin(\nu_{k^{3}} \tau)   
\hat V \hat {\bar \rho}_{k^1,k^2,k^{3}}^{\,[3:3]}(\tau).
\end{eqnarray}

From the definition, the initial conditions are set as $\hat {\bar \rho}^{[0:0]}(0)= \bf I$, where $\bf I$ is the unit operator, with all other hierarchy elements set to zero. The calculated elements $\hat {\bar \rho}_{k^1,\dots,k^{m}}^{\,[m:l]}(\beta \hbar)$ must be normalized after the integration over imaginary time is carried out by dividing by $Z_A =tr_A \{ \hat {\bar \rho}^{[0:0]}(\beta \hbar)\}$. A significant difference between the real-time HEOM, given in Eq.\eqref{eq:HEOM}, and the above imaginary-time HEOM is that the former contain damping terms proportional to $\gamma$ and $\nu_k$, whereas the latter contain sinusoidal terms. 
The imaginary-time HEOM readily yield the desired quantities, as they are solved by integrating over the pre-determined interval from $\tau=0$ to $\tau=\beta \hbar$, in contrast to the situation for the real-time HEOM, in which the integration must be carried out until convergence to the steady state is realized.
Any equilibrium expectation value of the system can be easily evaluated from $\hat {\bar \rho}^{[0:0]}(\beta \hbar)$. Moreover, we can evaluate the imaginary-time correlation functions\cite{CaoJCP1994} from Eq. \eqref{eq:ImHEOM} in the same manner that the real-time correlation functions are evaluated from the real-time HEOM.\cite{TanimuraJPSJ06} 

The correlated initial states for the real-time HEOM can be constructed from the hierarchy elements of the imaginary-time HEOM. The relations between the real-time and imaginary-time HEOM elements are similar to the relations between the expectation value of the collective bath oscillator coordinate and the real-time HEOM elements.\cite{Shi2012} Here, we present the relations between the two sets of elements up to second order in the system-bath interaction:
\begin{eqnarray}
\hat \rho_{0,\dots,0}^{(0)} (0)
= \frac{1}{Z_{A}} \hat {\bar \rho}^{\,[0:0]}  (\beta\hbar) , 
\label{eq:rho000}
\end{eqnarray}
\begin{eqnarray}
\hat \rho_{0,\dots,0}^{(1)} (0)
= -\frac{1}{Z_A} \left[ c_0''\hat {\bar \rho}_{0}^{\,[1:0]}  (\beta\hbar) +\sum_{k=1}^{K'}  
\left(\textcolor{red}{c_k''}\hat {\bar \rho}_{k}^{\,[1:0]}  (\beta\hbar) \textcolor{red}{+i c_k'} \hat {\bar \rho}_{k}^{\,[1:1]}  (\beta\hbar)  \right) \right],
\label{eq:rho100}
\end{eqnarray}
\begin{eqnarray}
\hat \rho_{0,\dots,j_k=1,0,\dots,0}^{(0)} (0)
= -\frac{1}{Z_A} \left( c_k' \hat {\bar \rho}_{k}^{\,[1:0]}  (\beta\hbar)  \textcolor{red}{-i c_k'} \hat {\bar \rho}_{k}^{\,[1:1]}  (\beta\hbar)  \right),
\label{eq:rho0k0}
\end{eqnarray}
\begin{eqnarray}
\hat \rho_{0,\dots,0}^{(2)} (0)
= \frac{1}{Z_A} \left[ c_0''^2 \hat {\bar \rho}_{00}^{\,[2:0]}  (\beta\hbar) + c_0''\sum_{k=1}^{K'}  
 \left(\textcolor{red}{c_k''} \hat {\bar \rho}_{0k}^{\,[2:0]}  (\beta\hbar) 
\textcolor{red}{+i c_k'}\hat {\bar \rho}_{0k}^{\,[2:1]}  (\beta\hbar)  \right) \right. \nonumber \\ \left. 
+\sum_{k=1}^{K'}  \sum_{k'=1}^{K'}  
\left(\textcolor{red}{c_k''c_{k'}''} \hat {\bar \rho}_{kk'}^{\,[2:0]}  (\beta\hbar)
\textcolor{red}{+2i c_k''c_{k'}'}\hat {\bar \rho}_{kk'}^{\,[2:1]}(\beta\hbar)
\textcolor{red}{- c_{k}' c_{k'}'}\hat {\bar \rho}_{kk'}^{\,[2:2]}(\beta\hbar) \right) 
  \right],
\end{eqnarray}
\begin{eqnarray}
\hat \rho_{0,\dots,j_k=1,0,\dots,j_{k'}=1,0,\dots,0}^{(0)}  (0)
&&= \frac{1}{Z_A} \left(c_k' c_{k'}'\hat {\bar \rho}_{k,k'}^{\,[2:0]}  (\beta\hbar) \textcolor{red}{-2 i c_k' c_{k'}'} \hat {\bar \rho}_{k,k'}^{\,[2:1]}  (\beta\hbar) \right. \nonumber \\
&& \left . 
\textcolor{red}{- c_k' c_{k'}'}\hat {\bar \rho}_{k,k'}^{\,[2:2]}  (\beta\hbar)   \right)  .
\label{rho011}
\end{eqnarray}
The elements $\hat \rho_{0,\dots,j_k=2,0,\dots,0}^{(0)} (0)$ are obtained by setting $k=k'$ in Eq.\eqref{rho011}. 
In practice, in order to evaluate the HEOM elements in the case of correlated initial conditions from the imaginary-time HEOM, the cutoff, ${K'}$, must be comparable to the cutoff of used for the real-time HEOM given in  Eq.\eqref{eq:HEOM}, $K$. 
If we only need the equilibrium distribution, Eq.\eqref{eq:rho000}, however, we may choose ${K'}$ even slightly smaller than $K/2$.

The equilibrium reduced density matrix has been evaluated from various approaches.\cite{CaoPRB2012, CaoJCP2012} Equation \eqref{eq:ImHEOM} allows us to calculate the same matrix elements using the reduced equation of motion approach. Moreover, this approach allows us to evaluate the correlated initial conditions of the real-time HEOM. 

\section{NUMERICAL RESULTS: CORRELATED INITIAL CONDITIONS, THERMODYNAMIC VARIABLES, AND EXPECTATION VALUES}

In this section, we report the results of numerical simulations that demonstrate the applicability and validity of the imaginary-time HEOM, given in Eq. \eqref{eq:ImHEOM}, for the system Hamiltonian
\begin{align}
\hat H_{A} &= \frac1{2} \omega_0 \hat \sigma_z + \frac1{2} \Delta \hat \sigma_x.
\label{eq:H_A}
\end{align}
We chose the system parameters as $\omega_0=1$ and $\Delta=0$ or $1$, and the bath parameters as $\beta=0.5 \sim 5$, $\eta=0 \sim 2$, and $\gamma=0.5$ for the system-bath interaction $\hat V= \hat \sigma_x$. We truncated the hierarchy by setting $\hat {\bar \rho}_{k^1,\dots,k^{{K'}+1}}^{\,[{K'}+1:l]} (\tau)=0$ for ${K'}=6$ in the imaginary-time HEOM, while we truncated by setting $\hat{\rho}_{j_1,\dots,j_K}^{(n)}(t)=0$ for $N\equiv n+\Sigma_{k=1}^K j_k =10$ with $K=7$ in the real-time HEOM. The 4th-order Runge-Kutta method was used for both the imaginary-time and real-time integrations with time steps of $\Delta \tau =1.0 \times 10^{-4}$ and $\Delta t =5.0 \times 10^{-4}$, respectively. The real-time HEOM were integrated from the factorized initial conditions $\rho_{0,\dots,0}^{(0)}(0)$ and $\hat{\rho}_{j_1,\dots,j_K}^{(n)}(0)=0$ at $t=0$, and steady states were realized between $t=100$ and $t=200$. 

\subsection{Correlated initial states}
First, we verified the accuracy of the imaginary-time HEOM by comparing the equilibrium state obtained from them with the steady state distributions obtained from the real-time HEOM for the temperatures $\beta=0.5, 1.0$, and $3.0$ with $\eta=1$ and $\Delta =1$. 
We found that the steady-state calculated from the real-time HEOM, $\hat \rho_{0,\dots,0}^{(0)}(0)$, deviates from the equilibrium state calculated from the imaginary-time HEOM, $\hat {\bar \rho}^{\,[0:0]} (\beta\hbar)/ Z_{A}$ by less than $0.0001\%$ of difference at $\beta=0.5$. The difference between the two results become larger for larger $\beta$ and for deeper hierarchy elements, because we solved the two kinds of HEOM using different truncation schemes. 
Other than this difference, however, the imaginary-time HEOM results are consistent with the real-time HEOM results. This also indicates that the steady-state elements obtained from Eq. \eqref{eq:HEOM} indeed represent the correlated thermal equilibrium state defined by Eq.\eqref{eq:correlatedHEOM}. 

Note that  
we must chose $K' \approx K$ in order to accurately calculate the real-time HEOM elements for the correlated initial conditions from the imaginary-time HEOM. Then, 
in order to obtain a better accuracy for deeper hierarchy elements in the imaginary-time HEOM, we used a small time step in the numerical integrations. 
For this reason, the computational costs for the real-time and imaginary-time HEOM were comparable. However, if we merely needed the equilibrium 
distribution $\hat {\bar \rho}^{\,[0:0]} (\beta \hbar)$  
to one percent accuracy, we could use a smaller cuttoff $K'$ and/or  
a larger time step for the imaginary-time HEOM and thereby reduce the computational costs  
to less than $1\%$ of that for the results reported here.

\begin{table}[tp] %
\begin{tabular}{|c|c|c|c|c|c|c|c|}\hline
$\beta \hbar$ & HEOM & $\; \;\rho^{(0)}_{0000000}\;\;$ & $\;\;\rho^{(1)}_{0000000}\;\;$ & $\rho^{(0)}_{1000000}$ & $\rho^{(0)}_{0010000}$ & $\rho^{(0)}_{0000010}$ & $\;\rho^{(2)}_{0000000}\;$  \\ 
\hline
0.5 
& imag & 0.617712 & 0.032478 & -0.000058 & -0.0000022 & -0.0000003 & 0.033128 \\  
& real & 0.617712 & 0.032478 & -0.000059 & -0.0000022 & -0.0000003 & 0.033116 \\ \hline
1.0 
& imag & 0.707858 & 0.062522 & -0.000395 & -0.0000157 & -0.0000020 & 0.033651 \\
& real & 0.707867 & 0.062530 & -0.000403 & -0.0000159 & -0.0000020 & 0.033562 \\ \hline
3.0 
& imag & 0.823431 & 0.121889 & -0.003975 & -0.0002132 & -0.0000280 & 0.031429 \\ 
& real & 0.823576 & 0.122238 & -0.004294 & -0.0002227 & -0.0000291 & 0.030249 \\ \hline
\end{tabular}
\caption{Hierarchy elements calculated from the imaginary-time HEOM and real-time HEOM for several values of the inverse temperature, $\beta$.}
\label{my-label}
\end{table} 

\subsection{Partition functions and thermodynamic variables}
Although with the real-time HEOM, we can calculate only the probability distribution, with the imaginary-time HEOM we are able to calculate thermodynamic variables via the partition function of the reduced system, $Z_A =tr_A \{ \hat {\bar \rho}^{[0:0]}(\beta \hbar)\}$. Note that the total partition function can be expressed as $Z_{tot}=Z_A Z_{B}$, where the partition function of the bath is given by
\begin{eqnarray}
Z_{B} =  \prod_j \frac{1}{2\sinh\left( \frac{\beta \hbar \omega_j }{2} \right)}.
\label{eq:Zbath}
\end{eqnarray}
Because we consider an infinite number of oscillators, however, the partition function of the bath cannot be determined. For this reason, we consider the system part, $Z_A$, only. We calculated the Helmholtz free energy, $F_A=-\ln(Z_A)/\beta$, the entropy, $S_A=k_B \beta^2 \partial F_A/ \partial \beta$, the internal energy, $U_A=-\partial \ln (Z_A)/ \partial \beta$, the heat capacity, $C_A=- k_B \beta^2 \partial U_A/\partial \beta$, and the susceptibility, $\chi_A=-(\partial F/\partial \Delta)$, from $Z_A$ for several values of $\beta$.
To obtain these quantities, we numerically integrated the imaginary-time HEOM for fixed $\eta=1$ and $\Delta=0$ to obtain $Z_A$ for $\beta$ satisfying $0.05 \ge \beta \ge 5$ at steps of $\Delta \beta =0.05$. For the susceptibility, we also calculated the free energy for $\Delta=0.05$ in order to evaluate the derivative with respect to $\Delta$ at $\Delta=0.025$.

The quantities mentioned above obtained using the imaginary-time HEOM are compared in Fig. 1 with the corresponding quantities for a system characterized by the canonical distribution, $Z_A^0=tr_A\{\exp[-\beta \hat H_A]\}$, with the same Hamiltonian $\hat H_A$ (with $\Delta=0$) and inverse temperature $\beta$. This corresponds to the partition function of the system in the case that the total partition function takes the factorized form
$Z_{tot}^0 = tr_A\{\exp[-\beta \hat H_A]\} tr_B\{ \exp[-\beta \hat H_B]\}$. 
The thermodynamic quantities are then given by 
$Z_A^0=2 \cosh(\beta \hbar \omega_0/2)$, $F_A^0 =-\ln(2 \cosh(\beta \hbar \omega_0/2))/\beta$,  $U_A^0 =-\tanh(\beta \hbar \omega_0/2)/2$,  
$S_A^0 =- k_B [(\beta \hbar \omega_0/2) \tanh(\beta \hbar \omega_0 /2)$ - $\ln(2 \cosh(\beta \hbar \omega_0/2))]$, and $C_A^0 =k_B (\beta \hbar \omega_0)^2/4 \cosh^2(\beta \hbar \omega_0/2)$. Also, note that the susceptibility for finite $\Delta$ is expressed as $\chi_A^0= \Delta\tanh \left[{\beta \hbar \sqrt{\omega _0^2 + \Delta^2)}/2} \right]/2$. The superscript "0" on these quantities indicates that these are calculated using the conventional statistical physics approach, which is equivalent to assuming a factorized thermal equilibrium state.

As seen in Fig. 2, in both cases of the spin-boson and factorized spin system, the entropy and internal energy decrease with the inverse temperature, while the heat capacities of both systems exhibit maxima at inverse temperatures near $\beta=2$, where the thermal excitation energy becomes comparable to the excitation energy. The entropy in the spin-boson case is larger than that in the factorized case at lower temperatures because the spin-boson system involves more degrees of freedom, due to the presence of the system-bath interaction. It is also seen that the internal energy is systematically lower in the spin-boson case than in the factorized case. This indicates that the bath absorbs some of the system energy through the interaction. The degree to which the system energy is absorbed by the bath increases as $\beta$ approaches the thermal excitation energy of the system and, as a result, the heat capacity of the spin-boson system becomes smaller than that of the factorized spin system near the peak position at $\beta=2$. Compared with the other thermodynamic variables, the difference between the susceptibilities in the two cases is small. This is because the system-bath interaction has the same form as the magnetic excitation, and the effects of $\Delta$ are suppressed by the strong system-bath interaction.

It is important to note here that those states regarded as the thermal equilibrium states in the two cases compared above are different. 
In the conventional treatment, the thermal equilibrium state of the system corresponds to the case of a factorized partition function, while in the present treatment of the spin-boson system, we consider the thermal equilibrium state of the total system.
Although the difference between the equilibrium thermodynamic quantities for the spin-boson system and the factorized spin system are rather minor in the static case considered in Fig. 2, the difference becomes significant when we study the dynamics of the system, because in this case, the positivity condition is often violated in the conventional treatment.
This may indicate that treatments based on the canonical distribution are inherently incompatible with dynamical states.
\begin{figure}
\begin{center}
\scalebox{0.7}{\includegraphics{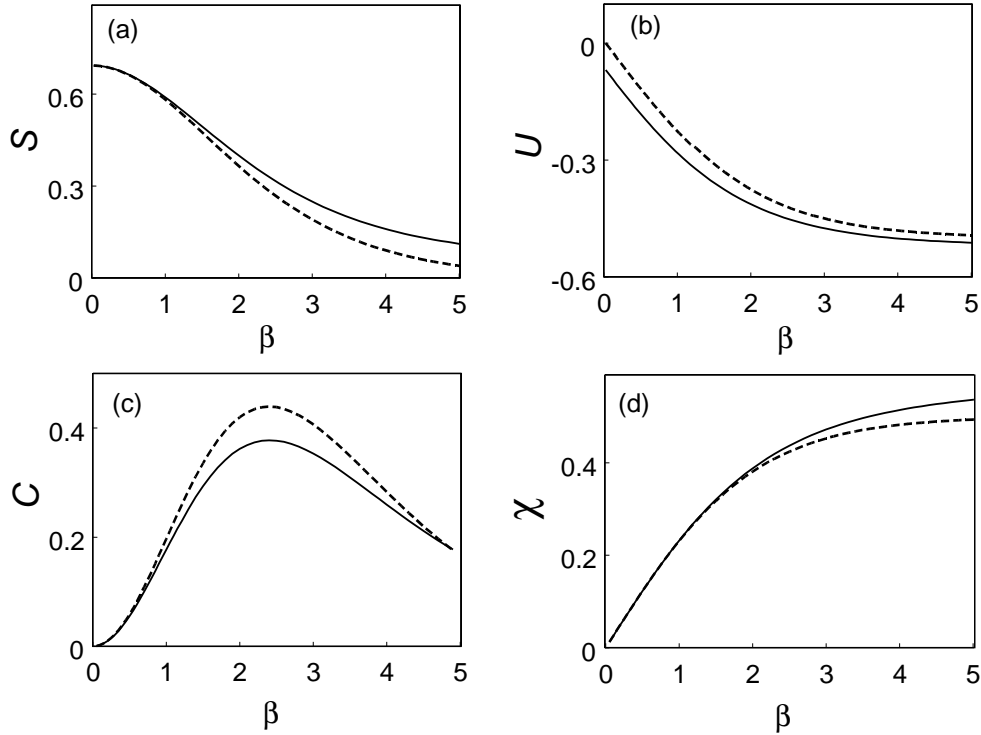}}
\end{center} 
\caption{\label{SUCCHI}The entropy, $S$, internal energy, $U$, heat capacity, $C$, and susceptibility, $\chi$, of a spin-boson system (solid curves) and a factorized spin system (dashed curves) as functions of the inverse temperature, $\beta$. The susceptibility, $\chi$, is calculated at $\Delta=\delta \ll \omega_0$, and is normalized by dividing by $\delta$.}
\end{figure}

\subsection{Auxiliary hierarchy elements and expectation values}
By utilizing the hierarchy elements, we can calculate expectation values of the system and bath. For example, the expectation value of the system energy, $\langle \hat H_A \rangle$, is obtained from Eq. \eqref{eq:rho000} as
\begin{eqnarray}
\langle \hat H_A \rangle=tr_A\{\hat H_A \hat \rho_{0,\dots,0}^{(0)} (0) \},
\end{eqnarray}
Using the first element of the hierarchy, the expectation value of the system-bath interaction, $\langle \hat H_I \rangle= tr\{\hat V \sum c_j \hat x_j \exp [-\beta \hat H_{tot} ]\}$, is evaluated as 
\begin{eqnarray}
\langle \hat H_I \rangle=tr_A\{\hat V \hat \rho_{0,\dots,0}^{(1)} (0) \}
+\sum_{k=1}^K tr_A \{\hat V  \hat  \rho_{0,\dots,j_k=1,0,\dots,0}^{(0)}  (0) \},
\end{eqnarray}
where $\hat \rho_{0,\dots,0}^{(1)} (0)$ and $ \rho_{0,\dots,j_k=1,0,\dots,0}^{(0)}  (0) $ are obtained from Eqs.\eqref{eq:rho100} and \eqref{eq:rho0k0}, respectively. 

In Fig. 2, we present the expectation values $\langle \hat H_A \rangle$ and  $\langle \hat H_I \rangle$ and the internal energy of the system, $U_A$, 
as obtained by numerically integrating Eq. \eqref{eq:ImHEOM} for various coupling strengths, $\eta$, at $\beta=1$ and $\beta=3$, with a step size of $\delta \eta=0.2$. At those temperatures, the system part of the energy increases linearly, while the interaction part decreases linearly as a function of the coupling strength, $\eta$, but the rates of decrease and increase are smaller for lower temperatures, because the thermal activity of the bath is lower in this regime. We should mention that the internal energy, $U_A$, contains the system part of the interaction energy but not the bath part. The system part and bath part of the interaction energies are calculated as $(U_A -\langle \hat H_A \rangle)$ and $\langle \hat H_I \rangle -(U_A -\langle \hat H_A \rangle)$, respectively. The internal energy decreases as a function of $\eta$ because the system part of the interaction energy also decreases as a function of $\eta$. Both the bath and system parts of the interaction energy decrease as function of $\eta$, but the bath part of the contribution is much larger than the system part, because the bath contains many degrees of freedom.

\begin{figure}
\begin{center}
\scalebox{0.7}{\includegraphics{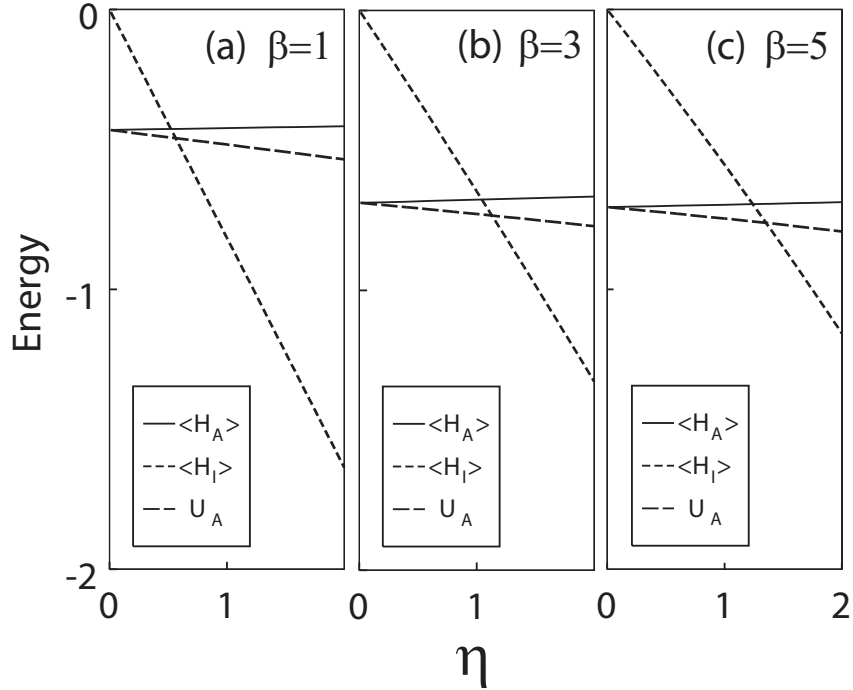}}
\end{center} 
\caption{\label{nu}The self energy of the system, $\langle \hat H_A \rangle$, the interaction energy, $\langle \hat H_I \rangle$, and the internal energy of the system, $U_A$, as function of the system-bath coupling strength, $\eta$, at (a) $\beta=1$, (b) $3$, and (c) $5$.}
\end{figure}

\section{CONCLUDING REMARKS}

In this paper, we derived the real-time and imaginary-time HEOM starting from the influence functional formalism with a correlated thermal initial state. It was shown that the thermal equilibrium state calculated from the imaginary-time HEOM is equivalent to the steady state solution of the real-time HEOM. 
Because the imaginary-time HEOM is defined in terms of integrals carried out over the definite time interval from $\tau'=0$ to $\tau'=\beta$ and because the elements of the imaginary-time HEOM are real, we were able to calculate the hierarchy elements more easily in this case than in the case of the real-time HEOM. 
Moreover, using the imaginary-time HEOM, we were able to calculate the partition function, and from this, we could directly obtain several thermodynamic quantities, namely, the free energy, entropy, internal energy, heat capacity, and susceptibility of the system in the dissipative environment. The expectation values of not only the system energy but also the system-bath interaction energy were also evaluated from the hierarchy elements obtained from the real- and imaginary-time HEOM. We found that for the purpose of studying equilibrium properties, rather than dynamical behavior, the imaginary-time HEOM is more usueful than the real-time HEOM. 

In this paper, we derived the HEOM for a system in the energy eigenstate representation, but extension to a system in the coordinate space representation (or the Wigner representation) is straightforward.\cite{KatoJPCB13,SakuraiJPSJ13,SakuraiNJP14,YaoYaoJCP14,TanimuraPRA91,TanimuraJCP92,SteffenTanimura00,TanimuraSteffen00} This extension will be helpful for identifying the pure quantum effects, because we can compare the quantum results with the classical results obtained from the classical limit of the imaginary-time HEOM.\cite{SakuraiJPC11,KatoJPCB13,TanimuraJCP92}

\begin{acknowledgments}
The author is grateful to Professor Hermann Grabert, Professor Gerhard Stock, Professor Heinz-Peter Breuer, and Dr. Lothar M\"uchenber and their group members at the Freiburg Institute for Advanced Studies for their hospitality. Financial support from the Humboldt Foundation, the University of Freiburg, and a Grant-in-Aid for Scientific Research (A26248005) from the Japan Society for the Promotion of Science are acknowledged. 
\end{acknowledgments}

\appendix
\section{Derivation of correlated influence functional}
Because the extension to a many oscillator system is straightforward, we start from a single oscillator bath, described by the Hamiltonian
\begin{eqnarray}
\hat H =  \frac{\hat p^2 }{2m } + \frac{1}{2}m \omega^2 \hat x^2  - V(t) \hat x .
\label{eq:BoS}
\end{eqnarray}
We then consider the density matrix elements with the three source terms $V$, $\bar V$, and $V'$ defined by\cite{Tanimura93,OkumuraPRE96,TanimuraOkumuraJCP97,SuzukiTanimuraPRE99}
\begin{eqnarray}
\rho(x,x'; t, \beta\hbar; \bf V) &&=  \int dx_0 \int dx_0' \int _{x(0)=x_0}^{x(t)=x} D[x(\tau)]
e^{ \frac{i}{\hbar}\int_0^{t} d\tau \left[\frac12{m \dot {x}^2}- \frac12 m \omega^2 x^2 + V(\tau) x \right] }
\nonumber \\
&&\times \int _{\bar x(0)=x_0'}^{\bar x(\beta \hbar)=x_0} D[\bar x(\tau')] 
e^{-\frac1{\hbar} \int_0^{\beta \hbar} d\tau' \left[\frac12{m \dot {\bar x}^2}+
 \frac12 m \omega^2 \bar x^2 -  {\bar V}(\tau') \bar x \right] } \nonumber \\
&&\times \int_{x'(0)=x_0'}^{x'(t)=x'} D[x'(\tau )] 
e^{- \frac{i}{\hbar}\int_0^{t} d\tau \left[\frac12{m {\dot x}'^2}- \frac12 m \omega^2 x'^2 + V'(\tau)x' \right] }. 
\label{eq:GV}
\end{eqnarray}
In order to evaluate $\rho(x,x'; t, \beta \hbar ; \bf V)$, we start from the Feynman propagator for Eq.\eqref{eq:BoS}, expressed as
\begin{eqnarray}
G(x,x_0,t; V)&&= \int _{x(0)=x_0}^{x(t)=x} D[x(\tau)]
e^{ \frac{i}{\hbar}\int_0^{t} d\tau \left[\frac12{m \dot {x}^2}- \frac12 m \omega^2 x^2 + V(\tau) x \right] }  \nonumber \\
&&=\sqrt{ \frac{m \omega}{2 \pi i  \hbar \sin (\omega t)} }
e^{\frac{i}{\hbar}S(x,x_0;t;V)},
\label{eq:Green}
\end{eqnarray}
where 
\begin{eqnarray}
S(x, x_0; t; V)&&= \frac{m \omega}{2 \sin (\omega t)} 
\left[ ({x}^2+{x_0}^2){\cos (\omega t) }- 2 xx_0 \nonumber \right. \\
&&+ \frac{2 x}{m \omega}\int_{0}^{t} dt'\, V(t') {\sin (\omega t')}  
+\frac{2 x_0}{m \omega}\int_{0}^{t} dt'\, V(t'){\sin \left(\omega (t-t') \right)}  \nonumber \\
&&\left.
-\frac{2}{m^2 \omega^2} \int_{0}^{t} dt'' \int_{0}^{t''} d t'\,
V(t''){\sin \left(\omega ( t- t'')\right)} V(t'){\sin (\omega  t')}
\right].
\label{eq:S_harm}
\end{eqnarray}
The equilibrium distribution, $\rho^{eq}(x,x'; \beta\hbar; {\bar V} )$, is obtained from Eq.\eqref{eq:Green} by replacing $i\tau/\hbar$ with $\tau'$. This yields 
\begin{eqnarray}
\rho^{eq}(x,x'; \beta\hbar; {\bar V}) &&= \int _{\bar x(0)=x'}^{\bar x(\beta \hbar)=x} D[\bar x(\tau')] 
e^{-\frac1{\hbar} \int_0^{\beta \hbar} d\tau' \left[\frac12{m \dot {\bar x}^2}+
 \frac12 m \omega^2 \bar x^2 -  {\bar V}(\tau') \bar x \right] } \nonumber \\
&&= \frac1{2\sinh\left(\frac{\beta \hbar \omega}2 \right) } \sqrt{\frac1{2 \pi \langle x^2\rangle}} 
\exp \left[ - \frac1{2 \langle x^2 \rangle} \left( \frac{x+x'}{2} - \bar r[{\bar V}; \beta \hbar] \right)^2 \right. \nonumber \\
&&- \left. \frac1{2\hbar^2} \langle p^2 \rangle (x-x')^2 + \frac{i}{\hbar} 
\bar p[{\bar V}; \beta \hbar] (x-x') +{\bar \Phi}[{\bar V}; \beta \hbar] 
\right], 
\label{eq:rhoqqt}
\end{eqnarray}
where
\begin{eqnarray}
\langle x^2\rangle =  \frac{\hbar}{2 m \omega} \coth \frac{\beta \hbar \omega}{2},
\label{eq:<q2B>}
\end{eqnarray}
\begin{eqnarray}
\langle p^2\rangle = \frac{\hbar m \omega}{2} \coth \frac{\beta \hbar \omega}{2},
\label{eq:<p2B>}
\end{eqnarray}
\begin{eqnarray}
\bar r[{\bar V}; \beta \hbar] 
=  \frac{1}{\hbar } \int_{0}^{\beta\hbar} d \tau'  {\bar V} (\tau') \bar L (\tau' ),
\label{eq:<barr>}
\end{eqnarray}
\begin{eqnarray}
\bar p[{\bar V}; \beta \hbar]  =  \frac{i m}{  \hbar } \int_{0}^{\beta\hbar} d \tau' {\bar V} (\tau') \dot {\bar L} (\tau' ),
\label{eq:<barp>}
\end{eqnarray}
and
\begin{eqnarray}
{\bar \Phi}[{\bar V}; \beta \hbar] =  \frac1{\hbar^2} 
\int_{0}^{\beta\hbar} d \tau'' \int_{0}^{\tau''} d \tau'  {\bar V} (\tau'') {\bar V} (\tau') \bar L (\tau'' - \tau' ).
\label{eq:PhitA2}
\end{eqnarray}
Here, we have
\begin{eqnarray}
\bar L(\tau') 
=  \frac{\hbar}{2m\omega} \frac{\cosh\left( \frac{\beta \hbar\omega}2 - \omega \tau' \right)}
{\sinh\left( \frac{\beta \hbar\omega}2 \right)}.
\label{eq:P1}
\end{eqnarray}
Note that the partition function for the oscillator itself, $Z\equiv \int dx \rho^{eq}(x,x; \beta\hbar;{\bar V}=0)$,  
can be obtained from Eq.\eqref{eq:rhoqqt} as
\begin{eqnarray}
Z = \frac{1}{2 \sinh\left( \frac{\beta \hbar \omega}{2} \right)}.
\label{eq:Zosc}
\end{eqnarray}

Using the counter path, we can express the total density matrix, Eq.\eqref{eq:GV}, as\cite{OkumuraPRE96,TanimuraOkumuraJCP97,SuzukiTanimuraPRE99}
\begin{eqnarray}
\rho(x,x'; t, \beta\hbar; \tilde V_C)&=& \int  D[\tilde x(s)]
e^{-\frac{i}{\hbar} \int_C ds \left[ \frac12{m \dot {\tilde x}^2(s)}- \frac12 m \omega^2\tilde x^2(s) + \tilde V_C(s)\tilde x(s) \right]} ,
\label{eq:GJ33}
\end{eqnarray}
where $\int D[\tilde {x}(s)]\equiv \int D[x(\tau)] \int D[\bar x(\tau')] \int D[x'(\tau)]$ and 
the contour paths are defined by Eqs.\eqref{eq:counterpath} and \eqref{eq:counterV}. We can obtain the full density matrix elements by simply replacing the integral $\int d \tau'$ in Eqs.\eqref{eq:rhoqqt}-\eqref{eq:PhitA2} with the contour integral $\int_C ds$. 
In the Wigner representation, we have the distribution
\begin{eqnarray}
W(p,r;t)=\frac1{2\pi \hbar} \int_{-\infty}^{\infty} e^{-i p q/\hbar} \rho(r+q/2,r-q/2;t ) dq.
 \label{eq:wigner}
\end{eqnarray}
After the normalization, this can be expressed as\cite{Tanimura93}
\begin{eqnarray}
W(&p&,r; t; \tilde V_C)= \frac1{2\pi} \sqrt{\frac1{ \langle p^2\rangle  \langle  x^2\rangle }} \nonumber \\
&&\times \exp \left[ - \frac1{2 \langle x^2\rangle } \left(r - \tilde r[\tilde V_C; t, \beta\hbar] \right)^2
- \frac1{2  \langle p^2\rangle }(p-\tilde p[\tilde V_C; t, \beta\hbar])^2 + {\tilde \Phi}[\tilde V_C; t, \beta\hbar] \right], 
\label{eq:wignerh}
\end{eqnarray}
where
\begin{eqnarray}
\tilde r[\tilde V_C; t, \beta\hbar] 
= -\frac{i}{\hbar} \int_{C} ds' \tilde V_C(s') L(s'),
\end{eqnarray}
\begin{eqnarray}
\tilde p[\tilde V_C; t, \beta\hbar] =  \frac{i m}{\hbar} \int_{C} ds' \tilde V_C(s') {\dot L}(s'),
\end{eqnarray}
\begin{eqnarray}
{\tilde \Phi}[\tilde V_C; t, \beta\hbar ] &=& - \frac1{\hbar^2} 
\int_{C} ds'' \int_{C'} ds' \tilde V_C (s'') \tilde V_{C'} (s') L(s''- s').
\label{eq:P2}
\end{eqnarray}
Here, $L(s)$ is the analytically continued Matsubara Green function obtained from Eq.\eqref{eq:P1} through the replacement $\tau' \rightarrow -i s'=-i(t'+i\tau')$, given by
\begin{eqnarray}
L(t'+i \tau' )=\frac{\hbar}{2m\omega} 
\frac{\cosh\left( \frac{\beta \hbar\omega}2 -\omega \tau' + i\omega t' \right)}
{\sinh\left( \frac{\beta \hbar\omega}2 \right)},
\label{eq:L0def}
\end{eqnarray}
and $C'$ represents the counter path for $s'$ that follows $s''$ along $C$ under the condition $s''> s'$. For the bath Hamiltonian appearing in Eq.\eqref{eq:bath} with the interaction $-V\sum \alpha_j x_j$, Eq.\eqref{eq:L0def} is expressed as Eq.\eqref{eq:Ltitau}. By tracing out $p$ and $r$, we obtain the influence functional for correlated initial conditions, $F[\tilde V_C; t, \beta\hbar]=\exp\left\{{\tilde \Phi}[\tilde V_C ; t, \beta\hbar]\right\}$.

\section{Influence phase}
For the counter path defined by Eqs.\eqref{eq:counterpath} and \eqref{eq:counterV}, the influence phase given in Eq.\eqref{eq:P2} is expressed as
\begin{eqnarray}
{\tilde \Phi}[{\bf V}; t, \beta\hbar ] 
&=&- \frac1{\hbar^2} \left[  \int_{0}^{t} dt''  \int_{0}^{t''} dt' V(t'') V(t') L(-[t''-t'])  \right. \nonumber \\
&&~~~~~ - i   \int_{0}^{t} dt'' \int_0^{\beta \hbar} d\tau' V(t'') {\bar V}(\tau')L(-t''-i\tau'+i\beta \hbar) \nonumber \\
&&~~~~~ - \int_{0}^{t} dt'' \int_0^{t} dt' V(t'')V' (t') L(t'-t''+i\beta \hbar) \nonumber \\
&&~~~~~ - \int_0^{\beta \hbar} d\tau'' \int_0^{\tau''} d\tau' {\bar V}(\tau'')  {\bar V}(\tau') L( i\tau''-i\tau') \nonumber \\
&&~~~~~ + i \int_0^{\beta \hbar} d\tau'' \int_0^t dt' {\bar V}(\tau'')  V' (t') L(t'+i\tau'') \nonumber \\
&&~~~~~\left.+  \int_0^t dt''\int_0^{t''} dt'  V' (t'') V' (t') L( -[t''-t'])
\right].
\label{eq:PhitBB} 
\end{eqnarray}
From the definitions $L(\pm t') \equiv \pm iL_1(t')+L_2(t')$ and Eqs.\eqref{eq:P1} and \eqref{eq:L0def}, we have the relations $L(i\tau') = \bar L(\tau')$,  $L(\pm t' + i \beta \hbar)= L(\mp t')$, and $L(t'+i\tau'+i\beta\hbar)=L(-t'-i\tau')$. With these, the influence phase can be expressed as Eq. \eqref{eq:PhitB}.

\end{document}